\documentclass{jetpl}
\twocolumn
\usepackage [cp1251]{inputenc}
\usepackage[english,russian]{babel}
\usepackage{color}
\rus

\title{Формирование долгоживущих сильнокоррелированных состояний света и поляризации среды при резонансном возбуждении полупроводниковых GaAs микрорезонаторов}

\rtitle{Формирование долгоживущих сильнокоррелированных состояний света и поляризации среды при резонансном возбуждении полупроводниковых GaAs микрорезонаторов}
\sodtitle{Формирование долгоживущих сильнокоррелированных состояний света и поляризации среды при резонансном возбуждении полупроводниковых GaAs микрорезонаторов}

\author{А.\,А.\,Деменев\/\thanks[1]{e-mail: demenev@issp.ac.ru},
С.\,И.\,Новиков, Д.\,Р.\,Домарецкий,
А.\,Л.\,Парахонский и
М.\,В.\,Лебедев
}

\rauthor{А.\,А.\,Деменев, С.\,И.\,Новиков, Д.\,Р.\,Домарецкий, А.\,Л.\,Парахонский и М.\,В.\,Лебедев}
\sodauthor{А.\,А.\,Деменев, С.\,И.\,Новиков, Д.\,Р.\,Домарецкий, А.\,Л.\,Парахонский и М.\,В.\,Лебедев}

\address{Институт физики твердого тела РАН, 142432 Черноголовка, Россия \\~\\}

\abstract{
При резонансной оптической накачке корреляционная функция интенсивностей излучения высокодобротного полупроводникового микрорезонатора имеет осциллирующий характер с неожиданно длинным периодом осцилляций и временем затухания, лежащими в наносекундном диапазоне. Показано, что природа осцилляций не связана со слабым Раби взаимодействием долгоживущих локализованных экситонных состояний в квантовой яме и электромагнитным полем моды микрорезонатора. На основе исследований отклика с высоким спектральным разрешением продемонстрировано, что необходимым условием возникновения осцилляций является совпадение положения и периода продольных лазерных мод с модуляционной структурой в спектре пропускания микрорезонатора.}


\begin{document}
\maketitle

\textbf{Введение}

Изучение неклассических состояний света привлекает в последнее время большое внимание в связи с многообещающими перспективами применения этих состояний в квантовой информатике. Примером неклассического состояния света, наблюдавшегося экспериментально, является флюоресцентное излучение одиночного атома, для которого корреляционная функция интенсивностей $g^{(2)}(0)=0$. Это явление было названо ``антигруппировкой'' фотонов \cite{Kimble77,Kimble78}. Для бифотонов -- фотонных пар, получаемых в процессе спонтанного параметрического рассеяния света в нелинейных кристаллах \cite{Mandel}, $g^{(2)}(0)$ может принимать значения превышающие 2, что принято называть ``сверхгруппировкой'' фотонов\cite{Kulik}. Фотоны как нельзя лучше подходят для передачи квантовой информации на большие расстояния из-за высокой скорости распространения и слабого взаимодействия с окружением, то есть слабой декогеренции. Существующие волоконно-оптические кабели связи, широко используемые в телекоммуникациях, вполне подходят также и для передачи квантовых состояний света, что имеет огромное значение для практической реализации различных квантовых протоколов. Развитие исследований в этой области направлено на решение двух главных задач – создание источников света, способных излучать по требованию заданное квантовое состояние и повышение яркости квантовых источников света. Обнаруженное нами недавно явление излучения неклассического света высокодобротным полупроводниковым микрорезонатором (МР) \cite{Lebedev} обладает достаточно высокими, по сравнению с другими источниками неклассического света, яркостью и временем корреляции. Природа и механизм возникновения неклассического света в полупроводниковом микрорезонаторе пока ещё до конца не ясны.

В работе \cite{Lebedev}, при измерении корреляционной функции интенсивностей света непрерывного полупроводникового лазера проходящего через МР в окрестности экситон - поляритонных резонансов наблюдалась сверхгруппировка фотонов в области нулевых задержек с последующими затухающими осцилляциями между состояниями антигруппировки и группировки фотонов.  Поляритонные состояния нашего МР имели времена жизни около 10 пс, в то время как период упомянутых выше осцилляций составлял величину масштаба 50 нс, а время их затухания могло достигать нескольких микросекунд. Мы провели подробное экспериментальное изучение условий возбуждения этих колебаний, результаты которого излагаются в настоящей работе.

\textbf{Образец и методика эксперимента}

Высокодобротный образец МР был выращен методом металлоорганической эпитаксии из газовой фазы и имеет верхнее (нижнее) брэгговское зеркало состоящее из 17(20) пар слоёв $Al_{0.13}Ga_{0.87}As/AlAs$ с оптической толщиной $\lambda/4$. Резонатор из GaAs длиной $3\lambda/2$ содержит шесть квантовых ям $In_{0.06}Ga_{0.94}As/GaAs$ толщиной 10 нм каждая. Расщепление Раби составляло $\Omega \approx 6$ мэВ. Добротность микрорезонатора составляла около 3000, что соответствует времени жизни фотона в резонаторе 3 пс. Представленные исследования выполнялись на образце и по схожей методике эксперимента как в нашей предыдущей работе \cite{Lebedev}, поэтому дополнительные сведения можно найти в \cite{Lebedev}.

Образец был смонтирован в оптическом гелиевом криостате с возможностью контролируемого изменения температуры. В настоящей работе представлены результаты, полученные при T=2.1K. Для возбуждения МР использовались два непрерывных перестраиваемых полупроводниковых лазера, генерировавшие многомодовый широкий спектр (5 и 2 нм) набора продольных мод. Излучение лазера направлялось вдоль нормали к поверхности образца и фокусировалось короткофокусной линзой на поверхность кристалла в пятно диаметром 50 мкм с угловой расходимостью 17 мрад. Дополнительная межзонная подсветка образца осуществлялась с помощью HeNe лазера с длиной волны излучения 632нм, также сфокусированного в пятно диаметром 50 мкм. Подсветка давала возможность быстро определить положения нижнего (LP) и верхнего (UP) поляритонных уровней в изучаемой области образца. Схему возбуждения поляритонной системы можно найти в \cite{Lebedev}. Излучение образца фокусировалось объективом на входной торец многомодового световода, который был подсоединен к двум кремниевым лавинным фотодиодам фирмы Perkin$\&$Elmer, работавшим в режиме счёта одиночных фотонов. Для обеспечения линейности отклика, значения скоростей счёта диодов поддерживались во всех экспериментах в диапазоне $10^{4}-10^{5}$ счётов в секунду. Корреляционные измерения проводились с использованием стандартной старт-стоп системы, включавшей два дискриминатора, преобразователь время-амплитуда и многоканальный анализатор. Максимальное временное разрешение нашей системы составляло около 300 пс.

Для изучения Раби осцилляций и релаксации в МР использовался перестраиваемый фемтосекундный титан-сапфировый лазер с длительностью импульса 100 фс. Частота следования импульсов составляла 3 кГц, исходная частота следования импульсов лазера 80 МГц была снижена до 3 кГц с помощью электрооптического затвора, пропускание которого в открытом состоянии превышало пропускание в закрытом на четыре порядка. Релаксация изучалась при помощи методики время коррелированного счета фотонов, когда преобразователь время-амплитуда (ТАС) запускался электрическим сигналом синхронизации от лазера, а останавливался сигналом с фотоприёмника, освещавшегося прошедшим через кристалл светом.

\begin{figure}
\begin{center}
\includegraphics[angle=0,width=1\linewidth]{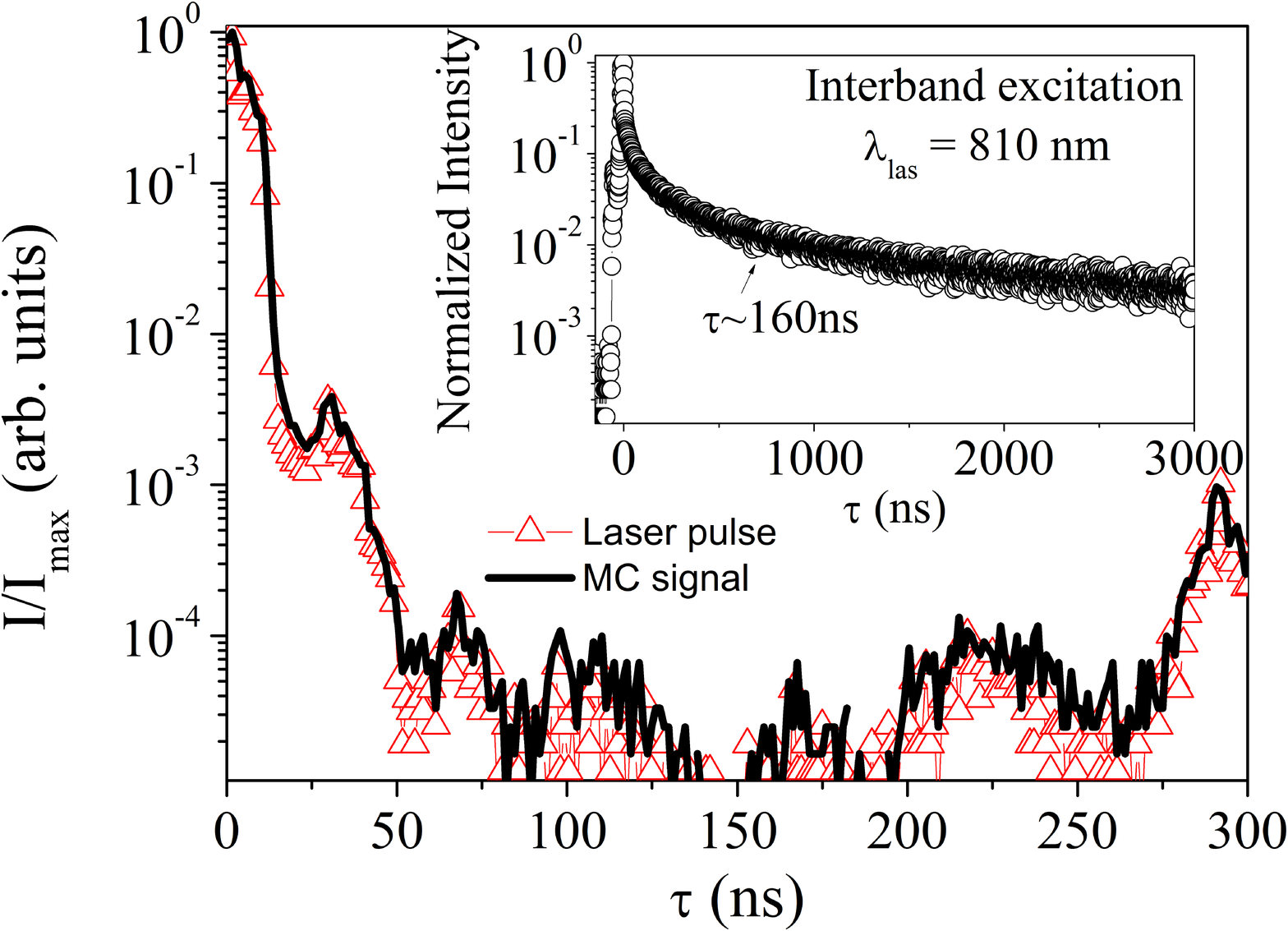}
\caption{Рис.1. Отклик МР на ультрокороткий (100фс) импульс титан-сапфирового лазера, полученный с помощью методики время коррелированного счета фотонов. Символами показан временной профиль импульса накачки, измеренный при тех же условиях. Вставка -- релаксационный отклик МР на нерезонансное импульсное возбуждение при $\lambda_{p}= 810$ нм. Т = 2.1 К.}
\end{center}
\end{figure}

Исследования сигнала из МР и спектра лазеров с высоким спектральным разрешением осуществлялось с помощью двойного монохроматора RAMANOR U1000, имеющего спектральное разрешение около 20 мкэВ. Для спектральной селекции лазерного излучения помимо названного монохроматора использовался двойной монохроматор МДР-6У с решётками 1200 мм$^{-1}$, работавший в режиме нулевой дисперсии.

\textbf{Результаты и обсуждение}

Целью наших экспериментов было выяснение условий, необходимых для генерации неклассического света в полупроводниковом микрорезонаторе. Это позволило бы, с одной стороны, пролить свет на физическую природу данного явления, а с другой – определить оптимальный режим возбуждения неклассического света. Вид корреляционной функции интенсивностей, получавшейся в наших экспериментах, хорошо совпадал с видом корреляционной функции полученной в теоретической работе \cite{Eleuch} для резонансного непрерывного возбуждения поляритонного состояния в полупроводниковом МР. Однако, характерные времена осцилляций в \cite{Eleuch} лежали в пикосекундном диапазоне, т.е. отличались от наблюдавшихся нами времен на четыре порядка. Также, затухание корреляционной функции в \cite{Eleuch} определялось затуханием внутрирезонаторного поля и экситонного уровня в квантовой яме (КЯ), что, опять же, имеет порядок десяти пикосекунд.

Особенности функции $g^{(2)}$ в \cite{Eleuch} связаны с осцилляциями Раби между экситонной поляризацией и светом при  резонансном взаимодействии света и вещества, усиленном наличием микрорезонатора , поэтому характерные времена осцилляций задаются силой экситон - фотонного взаимодействия и временем жизни фотона в МР. Наше первое предположение было, что наблюдаемые нами осцилляции также представляют собой осцилляции Раби между светом и долгоживущими локализованными экситонными состояниями в КЯ, взаимодействие которых со светом не усиливается, а подавляется наличием МР. Исходя из этого, мы ожидали увидеть длинновременные осцилляции Раби в релаксационном отклике МР при резонансном возбуждении фемтосекундными лазерными импульсами. Результаты этого эксперимента показаны на рис.1. Перестраиваемый фемтосекундный титан-сапфировый лазер с длительностью импульса 100 фс возбуждал микрорезонатор вдоль нормали к его поверхности на длине волны $\lambda_{p}= 855$ нм. Как видно из рис.1, мы не наблюдаем долгоживущих осцилляций интенсивности в релаксационном отклике, по крайней мере, на уровне до $10^{-3}$ от основного сигнала. Конечно, не исключено, что эти осцилляции просто ещё намного слабее и теряются в шумах, но такое предположение не согласуется с крайне высокой эффективностью возбуждения этих осцилляций в непрерывном режиме \cite{Lebedev}.

Если перестроить длину волны возбуждающего лазера выше по энергии фотона ($\lambda_{p}= 810$ нм), в область сильного поглощения квантовой ямы, то в релаксационном отклике хорошо видна долгоживущая люминесценция КЯ (вставка к рис.1) с временами релаксации порядка 100 нс. Отклик детектировался из области шириной 10нм вблизи основного экситонного уровня в КЯ. Таким образом, само существование долгоживущей поляризации среды в микрорезонаторе не вызывает сомнения.

\begin{figure}
\begin{center}
\includegraphics[angle=0,width=0.8\linewidth]{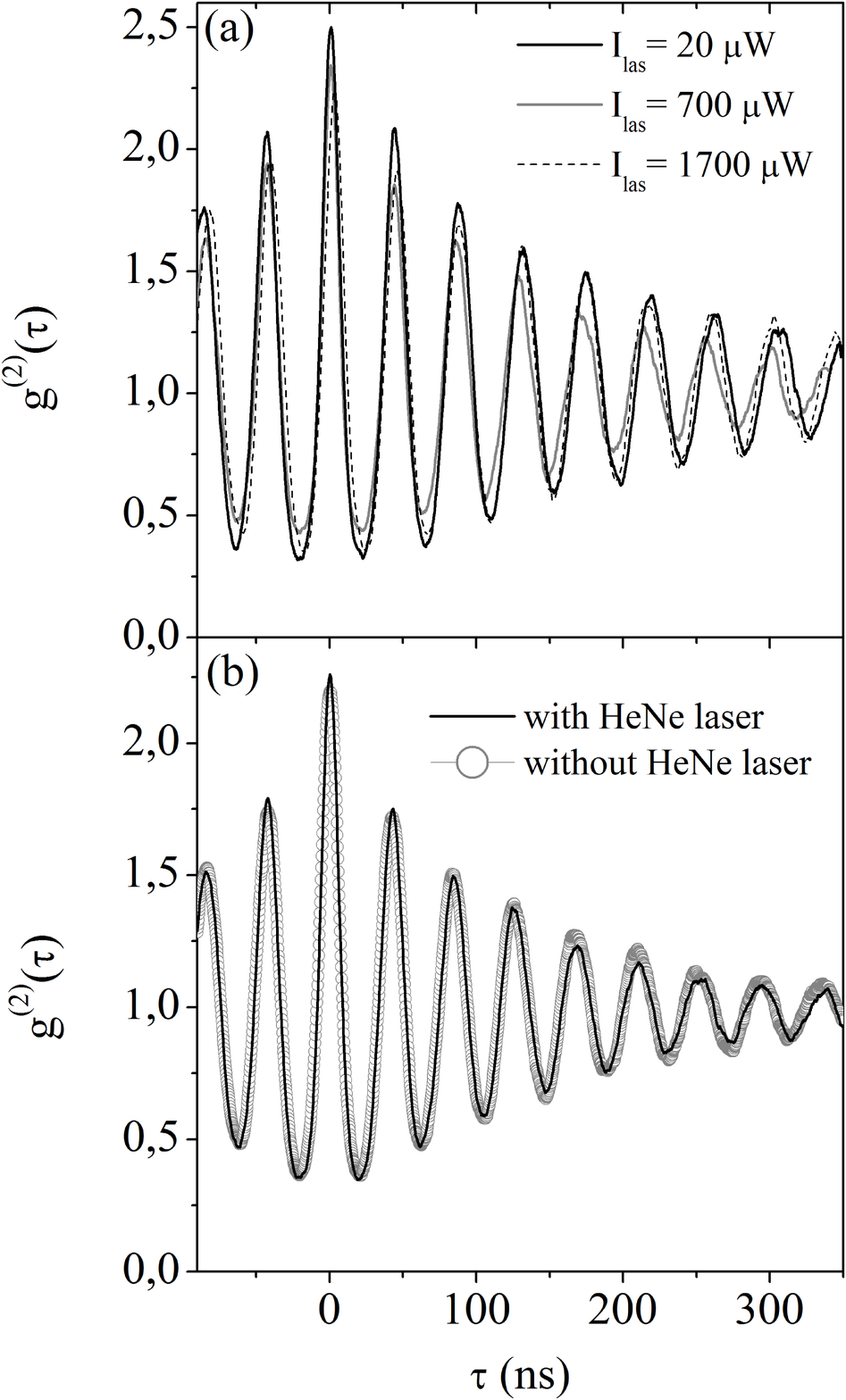}
\caption{Рис.2. (а)--Зависимость нормированной автокорреляционной функции интенсивностей отклика МР от мощности лазерной накачки; (b)--Влияние дополнительной межзонной подсветки гелий-неоновым лазером на временную динамику функции $g^{(2)}$. Т = 2.1 К.}
\end{center}
\end{figure}

Характерной чертой осцилляций Раби является корневая зависимость их частоты от интенсивности электромагнитного поля в резонаторе. С возрастанием интенсивности возбуждения МР, например, на порядок период наблюдаемых осцилляций должен был бы уменьшиться втрое. В эксперименте действительно наблюдается небольшое уменьшение периода осцилляций с повышением накачки, что показано на рис. 2(а). Однако, величина уменьшения гораздо меньше ожидаемого и при изменении мощности на два порядка составляет величину всего около 5 процентов. Это наблюдение не позволяет однозначно связать природу наблюдаемой автокорреляционной функции интенсивностей с осцилляциями Раби. С другой стороны, интерпретация в терминах осцилляций Раби остаётся возможной при дополнительном предположении о нелинейной зависимости поля внутри резонатора от поля накачки, так что сильное изменение внешнего поля приводит лишь к незначительному изменению поля внутри МР. Однако, поскольку эксперименты проводились для проверки теории из работы \cite{Eleuch}, где рассматривается случай слабого внутрирезонаторного поля (на уровне нескольких квантов света), то и мощности лазера выбирались из области линейного отклика МР на накачку.

Экситонная поляризация обычно весьма чувствительна к дополнительному возбуждению МР лазером в область сильного поглощения КЯ, что создаёт высокую плотность неравновесных носителей \cite{Demenev}. При этом, возможна нейтрализация заряженных примесей и, как следствие, изменение формы потенциала квантовой ямы, а также ускорение дефазировки поляризации из-за межчастичных взаимодействий. Влияние дополнительной подсветки гелий-неоновым лазером на корреляционную функцию интенсивностей сигнала пропускания полупроводникового лазера показано на рис.2(b). Несмотря на то, что мощность дополнительной подсветки достигала в наших экспериментах величины 1 мВт, что при диаметре пятна возбуждения 50 мкм обеспечивало создание плотности неравновесных носителей порядка $10^{7}$ см$^{-2}$ на одну КЯ, существенного влияния на генерацию неклассического света эта дополнительная подсветка не оказала.

Экспериментальные данные показывают, что неклассический свет возникает и тогда, когда частота возбуждающего лазера не совпадает ни с одним из поляритонных резонансов в нашей системе. Возникает вопрос: является ли наличие резонатора существенным для генерации неклассического света? Ответ на этот вопрос приведён на рис.3, где показан результат измерений автокорреляционной функции для двух областей МР. В первой области МР имеет цельные Брэгговские зеркала, во второй области одно из зеркал было стравлено наполовину методом химического травления. Спектры фотолюминесценции показывают, что добротность резонатора, при этом, существенно уменьшается (до $\sim 10^{3}$). Как следует из рисунка, видность функции $g^{(2)}$ для стравленной области заметно уменьшилась, однако период и время затухания осцилляций существенно не изменились. Неизменность периода осцилляций функции $g^{(2)}$ со значительным уменьшением добротности МР также подтверждает предположение, что природа явления не связана с Раби осцилляциями света и экситонных состояний в МР. Действительно, уменьшение добротности привело бы к расплыванию распределения светового поля в МР и, следовательно, к уменьшению силы осциллятора экситон-фотонного перехода. Таким образом, период осцилляций должен был бы заметно увеличится чего не наблюдается на рис.3.

\begin{figure}
\begin{center}
\includegraphics[angle=0,width=0.8\linewidth]{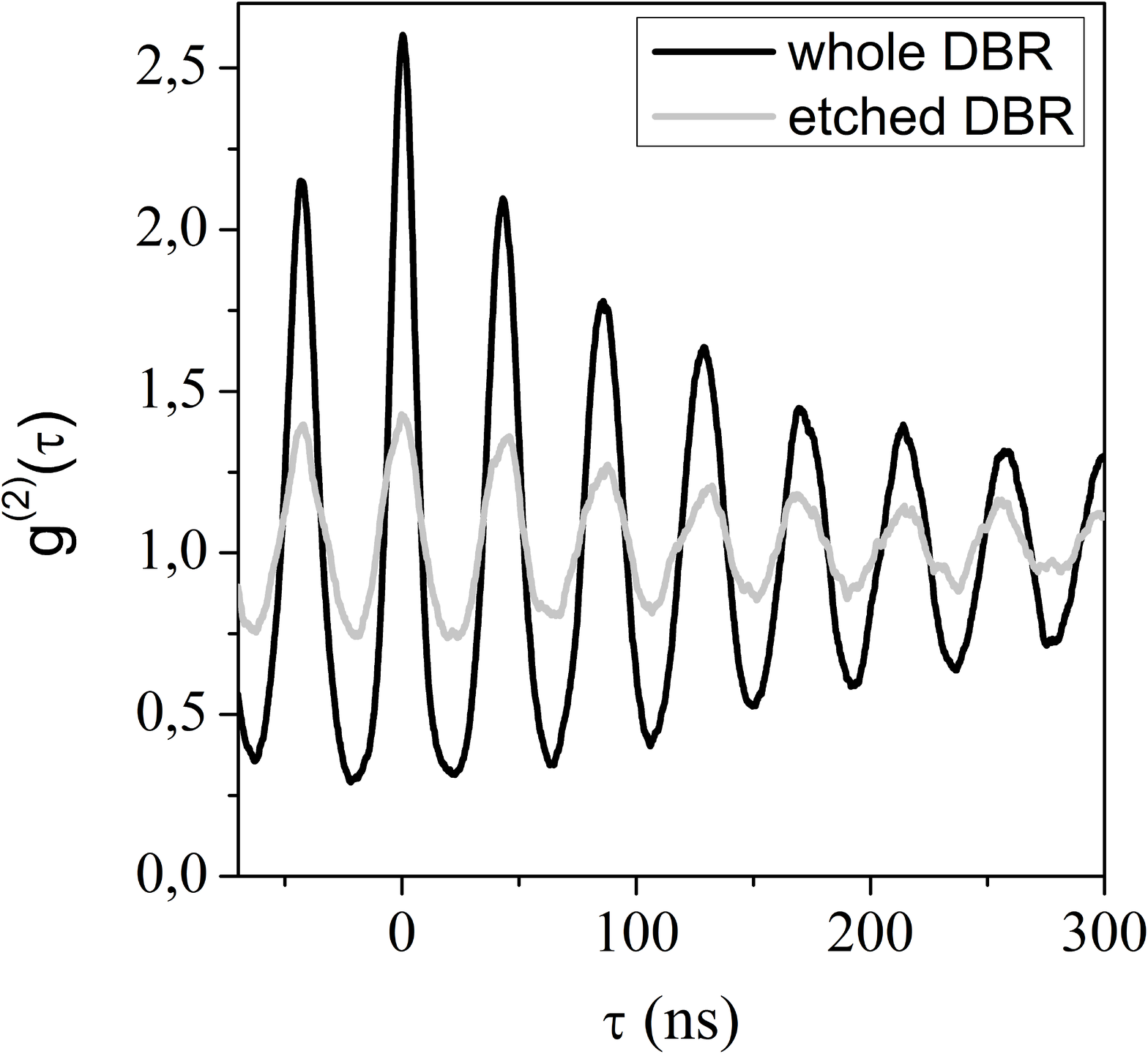}
\caption{Рис.3. Изменение характеристик нормированной автокорреляционной функции интенсивностей при понижении добротности резонатора. Т = 2.1 К.}
\end{center}
\end{figure}

Полупроводниковый лазер (пусть laser1), который использовался для возбуждения МР в \cite{Lebedev} имеет довольно широкий спектр генерации (5 нм), содержащий большое количество продольных мод. При подобном широком спектре возбуждения не вполне ясно, какие длины волн наиболее существенны для возбуждения неклассического света. Мы попробовали применить для возбуждения долгоживущих осцилляций непрерывный полупроводниковый лазер (пусть laser2), аналогичный применявшемуся ранее, но с меньшей шириной спектра генерации (2 нм). Несмотря на очень близкие условия возбуждения (мощность накачки, поляризация, центральная длина волны) корреляционная функция интенсивностей проходящего света демонстрировала только незначительный эффект группировки фотонов ($g^{(2)}(0)\approx1.11$), в то время как долгоживущие осцилляции полностью отсутствовали.

Подробное сравнение спектров излучения имевшихся в нашем распоряжении полупроводниковых лазеров между собой и со спектром пропускания МР, полученном в экспериментах по резонансному возбуждению микрорезонатора фемтосекундными импульсами титан-сапфирового лазера, выявило интересную особенность: интервал между соседними продольными модами полупроводникового лазера (laser1) эффективно возбуждающего неклассический свет с хорошей точностью совпал с периодом характерной модуляции спектра пропускания МР наблюдающейся в широком интервале энергий фотона (см. рис.4), а положения максимумов модуляционной структуры в спектре пропускания совпали с частотами мод лазера. В тоже время, для лазера (laser2) подобного согласования спектральной структуры не наблюдалось.

\begin{figure}
\begin{center}
\includegraphics[angle=0,width=0.9\linewidth]{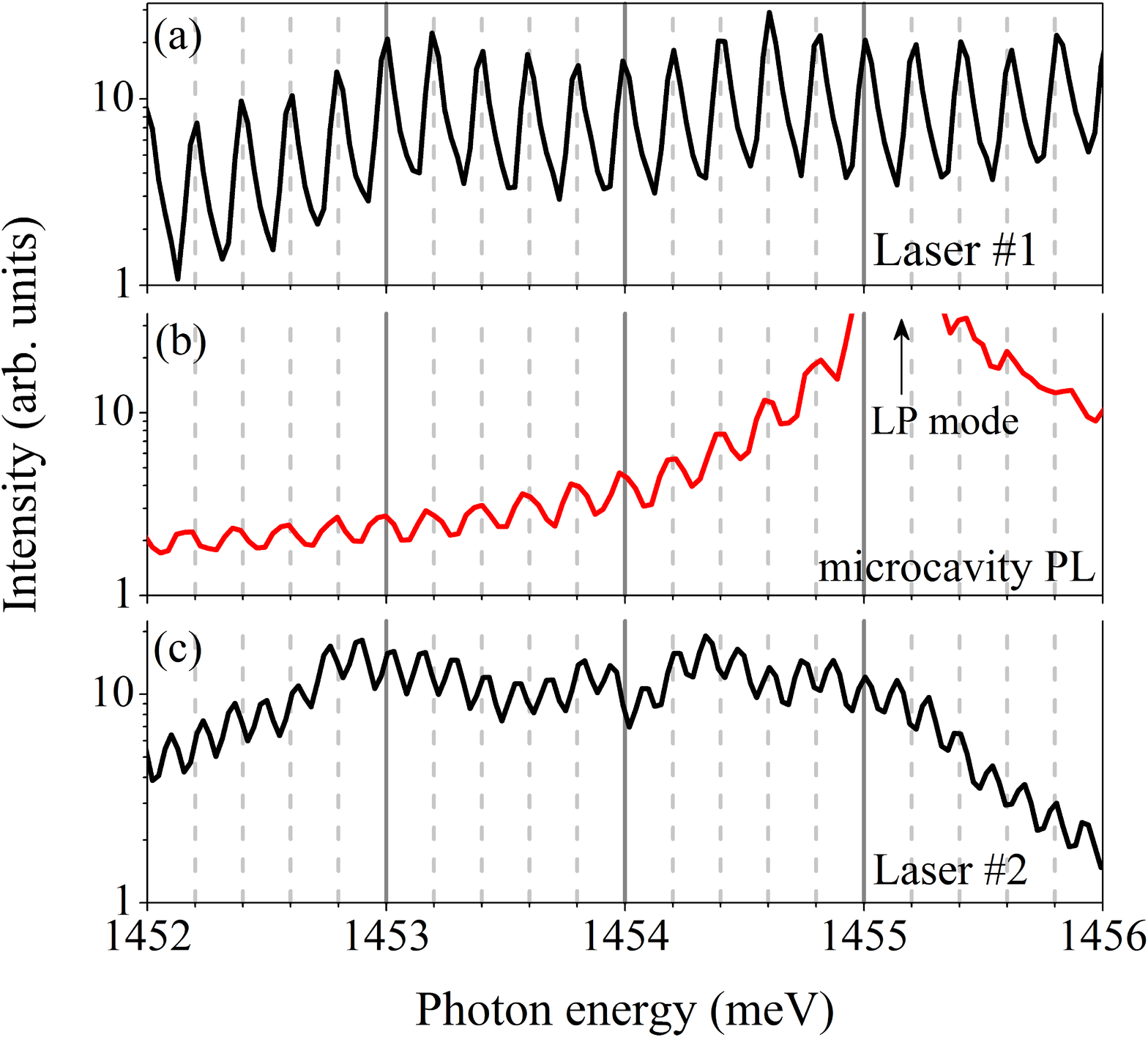}
\caption{Рис.4. Сравнение спектральных характеристик имевшихся в нашем распоряжении полупроводниковых лазеров между собой и со спектром пропускания МР. Т = 2.1 К.}
\end{center}
\end{figure}

Модуляция спектров излучения полупроводниковых МР, подобная показанной на рис.4(b), часто наблюдалась и ранее и, хотя природа её происхождения твёрдо не установлена, многие исследователи считают её своего рода интерференционным эффектом, возможно связанным с подложкой, на которой выращен МР. Такой взгляд представляется весьма логичным, поскольку вид этой модуляции очень похож на интерференционное сложение двух волн отражённых от граней тонкой плоскопараллельной пластинки. Пожалуй, единственной трудностью при таком объяснении является то, что модуляция наблюдается не в отраженном, а в проходящем свете и, следовательно, амплитуды двух интерферирующих волн должны были бы сильно различаться, в эксперименте же наблюдается очень существенная глубина модуляции, как если бы амплитуды этих волн были практически равны.

Большая ширина спектров накачивающих полупроводниковых лазеров позволяет осуществлять фильтрацию спектра возбуждения с помощью монохроматора. Мы сняли зависимость вида корреляционной функции интенсивностей от количества продольных мод, содержащихся в падающем на МР излучении лазера (laser1). Фильтрация для большого числа мод осуществлялась при помощи двойного монохроматора МДР-6У с решётками 1200 мм$^{-1}$, работавшем в режиме нулевой дисперсии. Для выделения одной или небольшого числа мод применялся монохроматор RAMANOR U1000. Зависимость видности корреляционной функции от числа продольных мод возбуждающего полупроводникового лазера показана на рис.5. С уменьшением числа мод наблюдается отчётливое увеличение видности. Важно, что неклассический свет наблюдается и при возбуждении одной единственной модой лазера. Отсюда следует, что возбуждение микрорезонатора несколькими частотами одновременно не является необходимым условием генерации неклассического света. Критичным условием такой генерации является, следовательно, совпадение частоты возбуждающего лазера с частотой, соответствующей максимуму модуляционной структуры в спектре пропускания МР.

\begin{figure}
\begin{center}
\includegraphics[angle=0,width=0.8\linewidth]{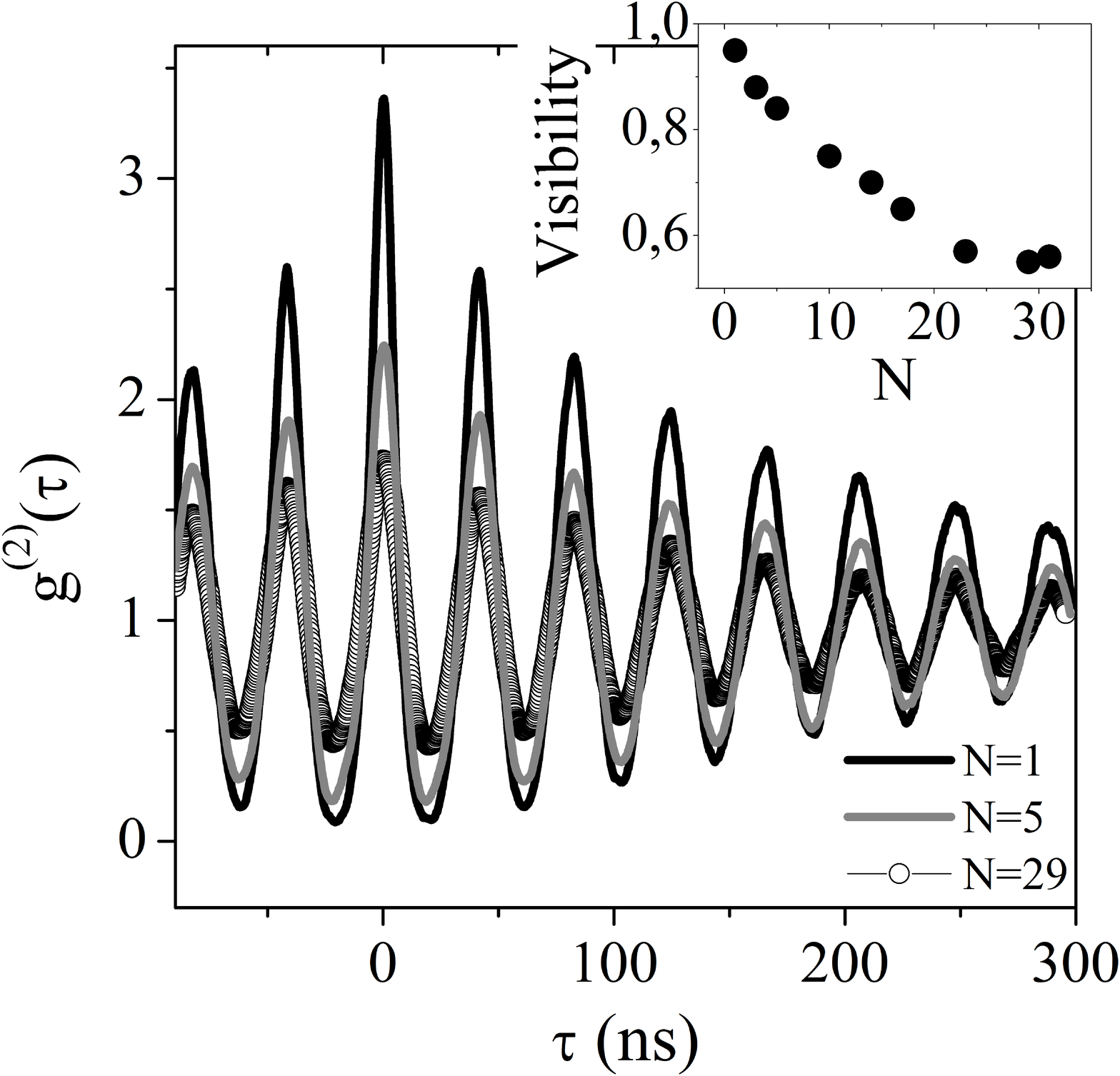}
\caption{Рис.5. Зависимость характеристик автокорреляционной функции интенсивностей пропускания от количества продольных мод, содержащихся в возбуждающем МР излучении лазера. Т = 2.1 К.}
\end{center}
\end{figure}

Совпадение частот возбуждающего лазера с частотами максимумов модуляционной структуры, необходимое для генерации неклассического света приводит, по видимому, к интерференционному эффекту сложения волн с близкими частотами. При подобном предположении появление нетипичного вида корреляционной функции интенсивностей можно качественно показать из следующих рассуждений. Пусть возбуждение МР происходит в непрерывном режиме лазером с одной продольной модой, тогда функцию $g^{(2)}$ можно записать через флуктуации интенсивности поля, $\delta I(t)=I(t)-I_{0}$:
\begin{equation}
{g^{(2)}(\tau) = \frac{\langle I(0) \cdot I(\tau) \rangle}{\langle I(0) \rangle \cdot \langle I(\tau)\rangle} = 1 + \frac{\langle \delta I(0) \cdot \delta I(\tau) \rangle}{(I_{0})^2},}
\end{equation} 
При сложении волн с частотами $\omega$ и $\omega+\delta$ в области детекторов получим, что $I(t) = I_{0} + 2*E_{1}*E_{2}*\cos(\delta t)$ и тогда для $g^{(2)}$ имеем выражение:
\begin{equation}
{g^{(2)}(\tau) = 1 + \frac{\langle (\delta I(0))^2 \rangle}{(I_{0})^2}\cdot \cos(\delta \tau),}
\end{equation} 
Заметим, что в выражение для среднего $\langle (\delta I(0))^2 \rangle$ входит амплитуда поля, например, $E_{2}$ связанная с поляризацией среды в МР. Как было показано выше (рис.1), время релаксации МР системы из области экситона может иметь масштаб сотен наносекунд. Релаксация поляризации среды приведет к постепенной дефазировке световых полей и, следовательно, $g^{(2)}\rightarrow1$. Математически это иллюстрируется, если комплексную величину $E_{2}$ записать как $E_{2} = E_{C2}*e^{i\Phi}$. Тогда, по аналогии с работой \cite{Lax}, можно показать , что $\langle e^{i\Phi(t)} \rangle = e^{-Dt/2}$, и пренебрегая диффузией фазы в лазере накачки получим:
\begin{equation}
{g^{(2)}(\tau) = 1 + A\cdot e^{-D\tau/2} \cdot \cos(\delta \tau),}
\end{equation} 
где $A = \frac{\langle (\delta I(0))^2 \rangle}{(I_{0})^2}$ -- среднее от действительных амплитуд световых полей при $\tau=0$.

Выражение для $g^{(2)}$ несложным образом обобщается на случай $N$ продольных лазерных мод со случайными фазами. В этом случае, функция $g^{(2)}$ также имеет вид (3), причем амплитуда $A$ (или видность корреляционной функции) вообще говоря зависит от $N$. Если все лазерные моды независимы, тогда $A\sim1/N$, в эксперименте, однако, величина $А$ линейно уменьшается с ростом $N$. В случае одной моды $A\approx1$, т.е. $g^{(2)}(0)\approx 2$, в эксперименте же наблюдается более значительная ``сверхгруппировка'' фотонов, $g^{(2)}(0)$ может достигать 3.5.

На текущий момент не до конца ясна природа рассогласования частот $\delta$, равного, в широком интервале изменения внешних параметров системы ``МР+лазер'', $\delta \approx 22$МГц.

\textbf{Заключение}

Таким образом, в данной работе, для прояснения природы генерации МР неклассического света, исследовалась автокорреляционная функция интенсивностей пропускания, демонстрирующая необычно длинные характерные времена, в различных условиях внешней накачки. На основе ряда исследований отклика МР: на резонансную накачку ультракороткими импульсами титан-сапфирового лазера; исследования мощностной зависимости характеристик функции $g^{(2)}$ при накачке непрерывным полупродниковым лазером; изучении влияния дополнительной межзонной подсветки гелий-неоновым лазером на вид $g^{(2)}(\tau)$, а также, влияния добротности резонатора на характеристики корреляционной функции можно утверждать, что природа осцилляций не связана со слабым Раби взаимодействием долгоживущих локализованных экситонных состояний в КЯ и внутрирезонаторным электромагнитным полем.

Исследования сигнала пропускания МР и спектров накачивающих лазеров с высоким спектральным разрешением продемонстрировали, что необходимым условием возникновения осцилляций автокорреляционной функции является совпадение положения и периода продольных лазерных мод с модуляционной структурой в спектре пропускания микрорезонатора. При этом, видность корреляционной функции увеличивается при уменьшении числа продольных мод возбуждающего непрерывного полупроводникового лазера. Неклассическое поведение $g^{(2)}$ наблюдается и при возбуждении одной единственной модой лазера.

Причина специфической модуляции спектров пропускания МР на текущий момент до конца не ясна, однако, как видится авторам, эффект перераспределения светового поля МР, в результате которого наблюдается неклассическое поведение $g^{(2)}(\tau)$ имеет интерференционную природу.

\textbf{Благодарности}

Мы признательны профессору В. Д. Кулаковскому и С. С. Гаврилову за полезные замечания и плодотворные обсуждения. Работа была поддержана грантом Президента РФ, грант МК-7844.2016.2.

\end{document}